\RequirePackage{fix-cm}
\documentclass[twocolumn]{svjour3}          
\smartqed  
\usepackage{graphicx}
\usepackage{amsmath,amssymb}
\usepackage{color}
\usepackage{lineno}
\begin{document}

\title{
Dynamics between order and chaos in conceptual models of glacial cycles
}

\titlerunning{Dynamics between order and chaos in conceptual models of glacial cycles}        

\author{Takahito Mitsui         \and
        Kazuyuki Aihara 
}


\institute{T. Mitsui \at
              FIRST, Aihara Innovative Mathematical Modelling Project, JST, 4-6-1 Komaba, Meguro-ku, Tokyo 153-8505, Japan \\
              Institute of Industrial Science, University of Tokyo, 4-6-1 Komaba, Meguro-ku, Tokyo 153-8505, Japan \\
              Tel.: +81-3-5452-6370\\
              Fax: +81-3-5452-6371\\
              \email{mitsui@sat.t.u-tokyo.ac.jp}           
           \and
           K. Aihara \at
              Institute of Industrial Science, University of Tokyo, 4-6-1 Komaba, Meguro-ku, Tokyo 153-8505, Japan
}

\date{}

\maketitle

\begin{abstract}
The dynamics of glacial cycles is studied in terms of the dynamical systems theory.
We explore the dependence of the climate state on the phase of the astronomical forcing 
by examining five conceptual models of glacial cycles proposed in the literature.
The models can be expressed as quasiperiodically forced dynamical systems.
It is shown that four of them exhibit a strange nonchaotic attractor (SNA),
which is an intermediate regime between quasiperiodicity and chaos.
Then, the dependence of the climate state on the phase of the astronomical forcing is not given by smooth relations,
but constitutes a geometrically strange set.
Our result suggests that SNA is a candidate for representing the dynamics of glacial cycles, 
in addition to well-known quasiperiodicity and chaos. 

\end{abstract}

\section{Introduction}
The climate change in the Quaternary is characterized by alternations between cold (glacial) and warm (interglacial) periods,
which are called glacial-interglacial cycles or alternatively glacial cycles.
The aim of this article is to explore the dynamics of glacial cycles by examining conceptual models proposed in the literature. 

Astronomical theories of ice ages, as typified by the Milankovitch theory (1941), attempt to explain glacial cycles
based on the change in the incoming solar radiation (insolation) 
due to the long-term variations of the Earth's orbital parameters.
Assuming a constant solar output and a perfectly transparent atmosphere, the insolation $F$ at a given latitude and season
is a function of the {\it astro-insolation parameters}:
the eccentricity $e$ of the Earth's orbit, 
the obliquity $\varepsilon $ (the inclination of the equator on the ecliptic), 
and the climatic precession $e \sin \tilde \omega$, where $\tilde \omega $ is
the longitude of the perihelion measured from the moving vernal equinox (Berger 1978).
That is, $F=F(e,\,\varepsilon ,\,e\sin \tilde \omega )$.

A motion expressed by a superposition of periodic motions of two or more incommensurate frequencies is called a {\it quasiperiodic motion}.
In the studies of the Quaternary climate changes, the astro-insolation parameters are often assumed to be quasiperiodic (Berger 1978) as follows: 
\begin{eqnarray*}
&e=e_0+\sum E_i \cos (\lambda _i t+\phi _i),&\\
&\varepsilon =\varepsilon ^*+\sum A_i \cos (f_i t+\delta _i),&\\
&e \sin \tilde \omega =\sum P_i \sin (\alpha _i t+\zeta _i).&
\end{eqnarray*}
Then, the insolation $F$ is also quasiperiodic in time (see Eq.~(\ref{eq:ins})).
The insolation variation $F(t)$ is known as {\it astronomical forcing}.
The primary frequencies of astronomical forcing are approximately
$1/19$ kyr$^{-1}$ and $1/23$ kyr$^{-1}$ for the climatic precession, and 
approximately $1/41$ kyr$^{-1}$ for the obliquity.
The assumption of quasiperiodicity of astronomical forcing is considered to be approximately valid over the last several million years (Berger and Loutre 1991) 
but may not be valid beyond this period because of the intrinsic chaoticity of the planetary motion (Lasker 1989).
However, since our study focuses on the glacial cycles during the past $0.7$~Myr after the mid-Pleistocene transition (Clark et al. 2006),
we construct our theory on the assumption of quasiperiodic astronomical forcing.  

Hays et al. (1976) presented strong evidence for astronomical theories of ice ages.
They found the primary frequencies of astronomical forcing in the geological spectra of marine sediment cores.
However, the dominant frequency in geological spectra is approximately $1/100$ kyr$^{-1}$,
although this frequency component is negligible in the astronomical forcing.
This is referred to as the ``$100$~kyr problem.''

To understand the climate responses to astronomical forcing, 
several concepts related to the dynamical systems theory have been proposed in the literature.
The {\it linear response} is a simple framework to 
explain the $41$, $23$, and $19$~kyr periodicities in the geological spectra. 
However, the linear response cannot appropriately account for the $100$~kyr periodicity (Hays et al. 1976).
Ghil (1994) explained the appearance of the $100$~kyr periodicity as
a {\it nonlinear resonance} to the combination tone $1/109$~kyr$^{-1}$
between precessional frequencies $1/19$~kyr$^{-1}$ and $1/23$~kyr$^{-1}$.
Contrary to the linear resonance, the nonlinear resonance can occur even if the forcing frequencies are far from the internal frequency of the response system.
Benzi et al. (1982) proposed {\it stochastic resonance} as a mechanism of the $100$~kyr periodicity,
where the response to small external forcing is amplified by the effect of noise.
Tziperman et al. (2006) proposed that the timing of deglaciations is set by the astronomical forcing
via the {\it phase-locking} mechanism.
If a phase locking occurs, the condition $qf=p_1f_1+p_2f_2+\cdots +p_Nf_N$ is satisfied,
where $f$ is the response frequency, $f_1,\,f_2,\,...,\,f_N$ are forcing frequencies, and $p_1,\,p_2,\,...,\,p_N$, and $q(\neq 0)$ are integers (Romeiras et al. 1987).
De~Saedeleer et al. (2013) suggested {\it generalized synchronization} (GS) to describe 
the relation between the glacial cycles and the astronomical forcing.
GS means that there is a functional relation between the climate state and the state of the astronomical forcing.
They also showed that the functional relation may not be unique for a certain model.
However, the nature of the relation remains to be elucidated.

In this research, we analyze five conceptual models of glacial cycles proposed in the literature, 
and show that the dependence of the climate state on the phase of the astronomical forcing may not be given by smooth relations,
but constitutes some geometrically strange set known as a {\it strange nonchaotic attractor} (SNA) (Grebogi et al. 1984; Kaneko 1984). 

The remainder of this article is organized as follows. 
In Section 2, we review the notions of quasiperiodically forced systems and attractors.
In Section 3, five models of glacial cycles are explained.
In Section 4, we investigate attractors of the models numerically, 
and show possible relations between the climate state and the phase of the astronomical forcing.
In Section 5, we show dynamical properties of the attractors.
In Section 6, we mention an implication of the results. 
Section 7 concludes this article.
%
%

\section{Quasiperiodically forced systems and attractors}
\subsection{Quasiperiodically forced systems}
We consider the dynamics of glacial cycles in the framework of {\it quasiperiodically forced dynamical systems}.
Let $\mathbb{T}^N=\mathbb{R}^N/(2\pi \mathbb{Z})^N$ be an $N$-dimensional torus.
In general, quasiperiodically forced dynamical systems can be expressed as
\begin{equation}
\begin{split} 
\dot{\mathbf{\theta }}&=\mathbf{\omega },\ \ \ \ \ \ \ \ \theta \in \mathbb{T}^N, \\
\dot{\mathbf{x}}&=\mathbf{g}\mathbf{(x},\mathbf{\theta }), \ \mathbf{x}\in \mathbb{R}^M,
\end{split} \label{eq:qp}
\end{equation}
where $\mathbf{\theta }=(\theta _1, \theta _2, ..., \theta _N)^\mathrm{T}$ is the phase of the drive subsystem, $\mathbf{x}=(x _1, x _2, ..., x _M)^\mathrm{T}$ is the state of the response subsystem,  
$\mathbf{g}\mathbf{(x},\mathbf{\theta })$ is a periodic function in each phase $\theta _i$, 
and $\mathbf{\omega }=(\omega _1, \omega _2, ..., \omega _N)^\mathrm{T}$ is a vector of incommensurate frequencies such that $k_1\omega _1+k_2\omega _2+\dots +k_N\omega _N= 0$ does not hold for any set of integers, $k_1$, $k_2$, ..., $k_N$, except for 
the trivial solution $k_1=k_2=\dots =k_N= 0$.
In the astronomical theory of climate change, 
$\theta $ corresponds to the phase of the astronomical forcing $F(t)$, 
and $\mathbf{x}$ corresponds to the climate state, as described in Section~3.

\subsection{Attractors}
Like most conceptual models of glacial cycles, we assume the existence of {\it attractors} in systems (\ref{eq:qp}).
An attractor is a compact set with a neighborhood such that, for almost every initial condition in this neighborhood, 
the limit set of the orbit as time tends to infinity is the attractor (Grebogi et al. 1984).
The closure of set of initial conditions which approach an attractor as time tends to infinity is called the {\it basin of attraction} of the attractor. 
Note that the attractor of system~(\ref{eq:qp}) is a subset of $\mathbb{T}^N\times \mathbb{R}^M$.
Thus, the dependence of the climate state on the phase of the astronomical forcing is represented by this attractor, after a certain transient time.
In quasiperiodically forced systems (\ref{eq:qp}), the following types of attractors typically appear:
{\it $N$-frequency quasiperiodic attractors}, {\it $L$-frequency quasiperiodic attractors with $L>N$},
{\it Strange nonchaotic attractors} (SNAs), and {\it chaotic attractors}. 

Lyapunov exponents give the mean exponential rates of divergence (or convergence) of nearby orbits.
Although the Lyapunov exponents depend on initial conditions,
they have the same set of values for almost every initial condition, under general circumstances. 
Such initial conditions or orbits which give the same set of values of the Lyapunov exponents are called {\it typical}.
A chaotic attractor is an attractor for which typical orbits
on the attractor have a positive Lyapunov exponent (Grebogi et al. 1984). 
That is, almost all orbits on the attractor have sensitive dependence on initial conditions. 
In quasiperiodically forced systems (\ref{eq:qp}), an attractor is chaotic if
the maximal conditional Lyapunov exponent (CLE) $\lambda$ is positive for typical orbits on the attractor, and nonchaotic otherwise. 
The maximal CLE $\lambda $ is defined as
$$\lambda =\lim _{t\to \infty}\frac{1}{t}\ln \frac{|\delta \mathbf{x}(t)|}{|\delta \mathbf{x}(0)|},$$
where $\delta \mathbf{x}(0)$ is an infinitesimal displacement of the orbit from $\mathbf{x}(0)$.
The evolution of the infinitesimal displacement is given by
the variational equation $\dot {\delta \mathbf{x}}=D_{\mathbf{x}} \mathbf{g}(\mathbf{x},\theta )\delta \mathbf{x}$.

An $N$-frequency quasiperiodic attractor is an attracting $N$-dimensional torus on which the motion is quasiperiodic.
$N$-frequency quasiperiodic attractors are characterized by a negative maximal CLE, $\lambda <0$.

An $L$-frequency quasiperiodic attractor with $L>N$ is 
an attracting $L$-dimensional torus on which the motion is quasiperiodic.
$L$-frequency quasiperiodic attractors with $L>N$ are characterized by a maximal CLE of zero, $\lambda =0$. 

A strange nonchaotic attractor (SNA) is a strange attractor
for which typical orbits have nonpositive Lyapunov exponents (Grebogi et al. 1984).
The strange attractor is defined as an attractor which is not a finite set of points and is neither piecewise differentiable curve or surface, nor a volume bounded by a piecewise differentiable closed surface. 
SNAs are characterized by a negative maximal CLE, $\lambda <0$.
It is proven that some SNAs are represented by a closure of graphs of almost everywhere discontinuous functions 
(Keller 1996; Alsed\`a and Costa 2009).

SNAs lie in an intermediate regime between quasiperiodicity and chaos.
The power spectrum of an SNA is intermediate one between quasiperiodicity and chaos, 
namely, a singular continuous (fractal) spectrum (Feudel et al. 1996).
SNAs are multifractal objects in the sense that the capacity dimension is larger than the information dimension 
(Ding et al. 1989; Hunt and Ott 2001).    
For comprehensive reviews of SNAs, refer to Prasad et al. (2001) as well as Feudel et al. (2006).

In the remainder of this article, we will not encounter examples of $L$-frequency quasiperiodic attractors with $L>N$,
and sometimes $N$-frequency quasiperiodic attractors are simply referred to as quasiperiodic attractors.

\subsection{Examples of $N$-frequency quasiperiodic attractors and an SNA}
A quasiperiodically forced map is regarded as a stroboscopic map 
obtained by sampling the state of the continuous-time system (\ref{eq:qp}) at time intervals corresponding to one of the forcing frequencies.
Let us consider the following quasiperiodically forced map studied by Glendinning (2002): 
\begin{equation}
\begin{split}
\theta (n+1)&=\theta (n)+\omega \pmod{1}, \\
x(n+1)&=2\sigma (\alpha +\cos 2\pi \theta (n))\tanh x(n), \label{eq:gopy} 
\end{split}
\end{equation}
where $(\theta ,x)\in [0,1)\times \mathbb{R}$, $n\in \mathbb{Z}$, $\omega\in \mathbb{R}\setminus \mathbb{Q}$, and 
$\sigma$ and $\alpha$ are non-negative parameters.

Figure~1(a) shows the quasiperiodic attractors for $\sigma =1.2$, $\alpha =1.001$, and $\omega =(\sqrt{5}-1)/2$ (solid lines).
Each attractor is represented by a graph of a smooth function.
If we denote the attractor in the region $x>0$ by $x=\phi (\theta )$,
the attractor in the other region is $x=-\phi (\theta )$ because of the symmetry $x\to -x$. 
The line $x=0$ is an unstable invariant set, namely, a {\it repellor}. 
In this case, the repellor constitutes the basin boundary of two attractors.

Figure~1(b) shows the SNA for $\sigma =1.2$, $\alpha =0.8$, and $\omega =(\sqrt{5}-1)/2$.
The SNA is represented by the closure of two graphs 
$x= \varphi (\theta )$ and $x=-\varphi (\theta )$ having following properties (Glendinning 2002; Keller 1996): \\
(i) $\varphi (\theta )\geq 0$ for all $\theta \in [0,1)$; \\
(ii) the union of two graphs $x=\varphi (\theta )$ and $x=-\varphi (\theta )$\\ 
is invariant under the map;\\
(iii) typical conditional Lyapunov exponents on the graphs are negative;\\
(iv) $\varphi (\theta )$ is discontinuous at almost all $\theta \in [0,1)$;\\ 
(v) $\varphi (\theta )$ is upper semi-continuous;\\
(vi) $\varphi (\theta )=0$ on a dense, but measure zero, set of values of $\theta \in [0,1)$.\\
\begin{figure}[]
\begin{center}
\includegraphics[scale=1.0,angle=0]{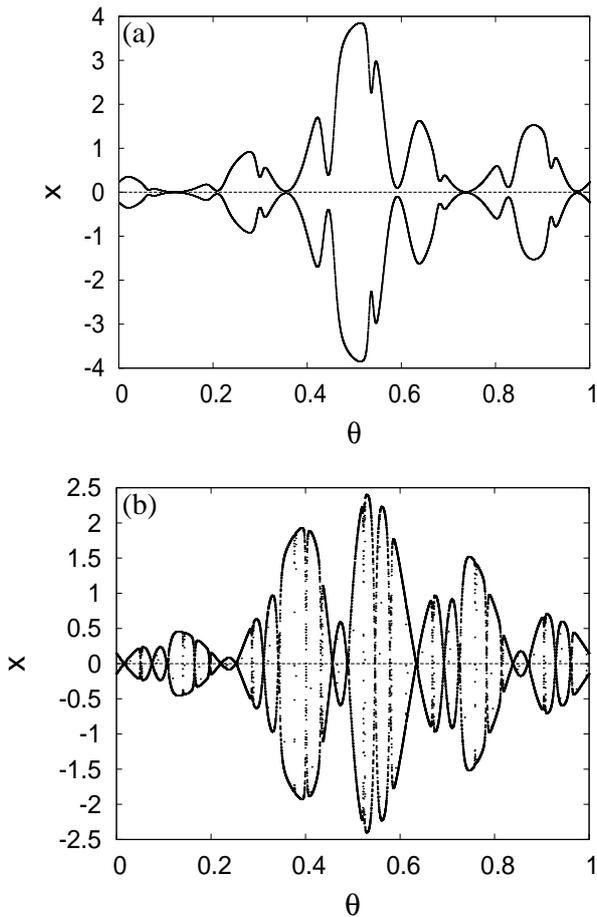}
\caption{Attractors in map~(\ref{eq:gopy}): {\bf a} Quasiperiodic attractors for $\sigma =1.2$, $\alpha =1.001$, and $\omega =(\sqrt{5}-1)/2$ ({\it upper} and {\it lower solid lines}). The maximal CLE is $\lambda =-0.3677$ for each attractor. 
{\bf b} SNA for $\sigma =1.2$, $\alpha =0.8$, and $\omega =(\sqrt{5}-1)/2$ ({\it dots}). The maximal CLE is $\lambda =-0.2124$.
In each panel, the {\it dashed line} $x=0$ is the unstable invariant set} \label{fig:gopy_attr}
\end{center}
\end{figure}

Unlike quasiperiodic attractors, SNAs contain some repellors in themselves (Stark 1997).
See the repellor $x=0$ embedded in the SNA in Fig.~\ref{fig:gopy_attr}(b).
Due to such repellors, the orbits on SNAs become temporarily unstable and sensitive to perturbations, as shown in Section 5.   

\subsection{Numerical characterization of strangeness of attractors}
SNAs and $N$-frequency quasiperiodic attractors can be numerically distinguished by the phase sensitivity exponent (Pikovsky and Feudel 1995) or by the parameter sensitivity exponent (Nishikawa and Kaneko 1996). 
First, let us consider the case of the continuous-time system (\ref{eq:qp}).
The phase sensitivity function $\Gamma (t)$ of variable $x_j$ with respect to variable $\theta _i$ is defined as 
$$
\Gamma (t)=\min _{\{\mathbf{x}(0),\theta (0)\}}\left\{ \max _{0\leq \tau \leq t} \left| \frac{\partial x_j(\tau )}{\partial \theta _i(0)}\right| \right\}.
$$
The derivative $\partial x_j(\tau )/\partial \theta _i(0)$ 
characterizes the sensitivity of $x_j$ with respect to a change of $\theta _i$.
This derivative is obtained by integrating 
\begin{equation}
\frac{d}{dt}\frac{\partial x_j}{\partial \theta _i(0)}=\sum _{k=1}^{M}\frac{\partial g_j(\mathbf{x},\mathbf{\theta })}{\partial x_k}\frac{\partial x_k}{\partial \theta _i(0)}+\frac{\partial g_j(\mathbf{x},\mathbf{\theta })}{\partial \theta _i(0)}, \label{eq:phs}
\end{equation}
along the solution ($\mathbf{\theta }(t),\mathbf{x}(t)$).
For SNAs, the phase sensitivity function grows as $\Gamma (t)\simeq t^{\mu}$ with $\mu >0$, 
where $\mu $ is called the {\it phase sensitivity exponent}.
On the other hand, for $N$-frequency quasiperiodic attractors, $\Gamma (t)$ is bounded and $\mu =0$.

Assume that function $\mathbf{g}(\mathbf{x},\theta )$ contains some parameter, say $\alpha $. 
The parameter sensitivity function $\Psi (t)$ of variable $x_i$ with respect to parameter $\alpha$ is defined as
$$\Psi (t)=\min _{\{\mathbf{x}(0),\theta (0)\}}\left\{ \max _{0\leq \tau \leq t} \left| \frac{\partial x_j(\tau )}{\partial \alpha}\right| \right\}.$$
The derivative $\partial x_j(\tau )/\partial \alpha$ 
characterizes the sensitivity of $x_j$ with respect to a change of $\alpha$.
This derivative is obtained by integrating 
\begin{equation}
\frac{d}{dt}\frac{\partial x_j}{\partial \alpha}=\sum _{k=1}^{M}\frac{\partial g_j(\mathbf{x},\mathbf{\theta })}{\partial x_k}\frac{\partial x_k}{\partial \alpha}+\frac{\partial g_j(\mathbf{x},\mathbf{\theta })}{\partial \alpha}, \label{eq:pas}
\end{equation}
along the solution ($\mathbf{\theta }(t),\mathbf{x}(t)$).
For SNAs, the phase sensitivity function grows as $\Psi (t)\simeq t^{\nu}$ with $\nu >0$, 
where $\nu $ is called the {\it parameter sensitivity exponent}.
On the other hand, for $N$-frequency quasiperiodic attractors, $\Psi (t)$ is bounded and $\nu =0$.

For discrete-time systems, such as Eq.~(\ref{eq:gopy}), the methods for characterizing the strangeness of attractors
are almost the same except that time evolutions corresponding to
Eqs.~(\ref{eq:phs}) and (\ref{eq:pas}) become discrete in time (Pikovsky and Feudel 1995; Nishikawa and Kaneko 1996).

\begin{figure}[]
\begin{center}
\includegraphics[scale=0.55,angle=0]{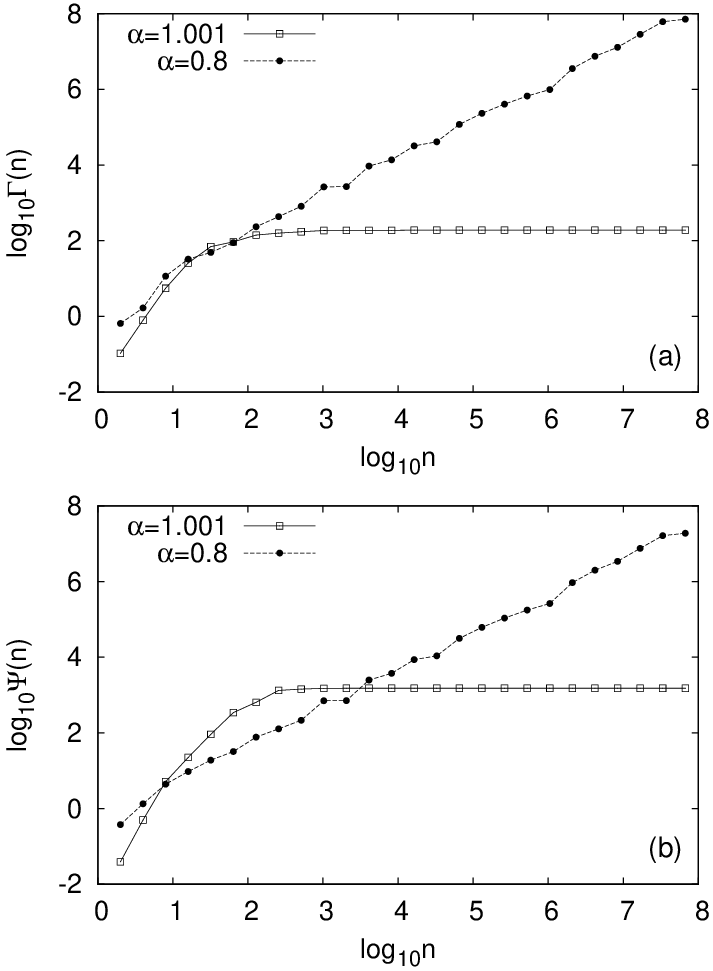}
\caption{{\bf a} Phase sensitivity function $\Gamma (n)$ for the quasiperiodic attractors of map~(\ref{eq:gopy}) with $\sigma =1.2$, $\alpha =1.001$, and $\omega =(\sqrt{5}-1)/2$ ({\it open square}) and for the SNA with $\sigma =1.2$, $\alpha =0.8$, and $\omega =(\sqrt{5}-1)/2$ ({\it filled circle}).
{\bf b} Paramater sensitivity function $\Psi (n)$ for the same attractors
} \label{fig:gopy_gamma}
\end{center}
\end{figure}

Figure~\ref{fig:gopy_gamma} shows the time evolutions of the phase sensitivity function $\Gamma (n)$ 
and the parameter sensitivity function $\Psi (n)$ with respect to $\alpha $ for the attractors in Fig.~\ref{fig:gopy_attr}.
The functions $\Gamma (n)$ and $\Psi  (n)$ saturate for large $n$ for the quasiperiodic attractors, 
but show power-law divergences with exponents $\mu =1.03$ and $\nu =0.99$ for the SNA.

\section{Models}
\subsection{Astronomical insolation forcing}
Milankovitch (1941) considered the high-latitude northern hemisphere summer insolation as a decisive factor for glaciations.
We use the insolation $F(t)$ at $65^\circ $N on the day of the summer solstice in the simulations of all the models.
For simplicity in computation, the insolation $F(t)$ is calculated using the following formula given in De~Saedeleer et al. (2013)\footnote{According to De~Saedeleer et al. (2013), the formula is valid over the past 1~Myr, and its mean error compared to Lasker et al. (2004) is 6.7 W/m$^2$ with peaks at 27.5 W/m$^2$.}: 
\begin{equation*} 
F(t)=\frac{1}{a}\sum _{i=1}^N [s_i\sin (\omega _it)+c_i\cos (\omega _it)],
\end{equation*}
where $N=35$ is the number of different frequencies $\omega _i$ included in the sum.
The values of frequencies $\omega _i$ are those in Berger (1978).
The coefficients $s_i$ and $c_i$ are obtained by linear regression 
on the insolation values calculated after Berger (1978) (De~Saedeleer et al. 2013). 
To make the paper self-contained, the values of $\omega _i$, $s_i$, and $c_i$ are listed in Table~\ref{tab:3} in Appendix~1.
The parameter $a$ is a scale factor, the value of which is different among models.
The top panel of Fig.~\ref{fig:compare_700} shows the insolation $F(t)$ normalized to have zero-mean and unit variance (with $a=23.58$ W/m$^2$).

Introducing variables $\theta _i(t)=\omega _it$ ($i=1,2,...,N$), the above equation for $F(t)$ can be written as 
\begin{equation}
\begin{split} 
F(t)&=\frac{1}{a}\sum _{i=1}^N [s_i\sin \theta _i(t)+c_i\cos \theta _i(t)], \label{eq:ins}\\
\dot \theta _i&=\omega _i, \ \ \  \theta _i(0)=0 \ \ \ (i=1,2,...,N).
\end{split}
\end{equation}
Note that $\theta _i$ are the phase variables of the astronomical forcing in climate models.

\subsection{Conceptual models of glacial cycles}
\subsubsection{Saltzman--Maasch model}
Saltzman and Maasch (1990) introduced a global model of the late Cenozoic climate.
We will refer to this model as SM90.
The model consists of differential equations for the global ice volume $x$,
the atmospheric CO$_2$ concentration $y$, and the global mean deep water temperature $z$,
where $x$, $y$, and $z$ are dimensionless and denote deviations from $1$~Myr averages.
Neglecting stochastic perturbations, the model is written in the following deterministic from: 
\begin{equation} 
\begin{split}
\tau \dot x& =-x-y-vz-uF(t),\\
\tau \dot y&=-pz+ry+sz^2-wyz-z^2y,\\
\tau \dot z&=-q(x+z),
\end{split} \label{eq:SM90}
\end{equation}
where $t$ is the time in kiloyears, and $\tau =10$~kyr. 
The parameters $p$, $q$, $r$, $s$, and $w$ are freely chosen, 
but the remaining parameters $u$ and $v$ are functions of the other parameters.
The insolation forcing $F(t)$ is given by Eq.~(\ref{eq:ins}) with $a=23.58$ W/m$^2$.

Saltzman and Maasch (1990) simulated glacial cycles for the last $500$ kyr 
by choosing the parameter values as set A in Table~1.
Hargreaves and Annan (2002) estimated the parameter values of SM90 as set B using a data assimilation technique.

\begin{table}
\caption{Sets of parameter values used for the SM90 and MS90 models}
\label{tab:1}       
\begin{tabular}{llllllll}
\hline\noalign{\smallskip}
Set & $p$ & $q$ & $r$ & $s$ & $u$ & $v$ & $w$  \\
\noalign{\smallskip}\hline\noalign{\smallskip} 
A & $1.0$ & $2.5$ & $0.9$ & $1.0$ & $0.6$ & $0.2$ & $0.5$ \\
B & $0.82$ & $4.24$ & $0.95$ & $0.53$ & $0.32$ & $0.02$ & $0.66$ \\
C   & $1.0$ & $1.2$ & $0.8$ & $0.8$ & $0.7$ & $0.0$ & $0.0$ \\
\noalign{\smallskip}\hline
\end{tabular}
\end{table}

\subsubsection{Maasch--Saltzman model}
Prior to the SM90 model, Maasch and Saltzman (1990) studied one of their previous models.
This older model, which we call the MS90 model, can be obtained from the SM90 model by setting $v=w=0$.
The other parameter values are chosen as set C in Table~\ref{tab:1}.

\subsubsection{Crucifix--De~Saedeleer--Wieczorek model}
One of the simplest systems that exhibit self-sustained oscillations is the van der Pol oscillator. 
With slight modifications and addition of the insolation forcing to the van der Pol oscillator, 
Crucifix (2012) introduced the following model:
\begin{equation} 
\begin{split}
\tau \dot x& =-[y+\beta +\gamma F(t)],\\
\tau \dot y&=-\alpha [y^3/3-y-x],
\end{split} \label{eq:CSW}
\end{equation}
where $x$ denotes the global ice volume, and
$y$ is a conceptual variable for realizing self-sustained oscillations. 
The insolation forcing $F(t)$ is given by Eq.~(\ref{eq:ins}) with $a=11.77$ W/m$^2$. 
For the unforced case $\gamma =0$, Eq.~(\ref{eq:CSW}) has a stable equilibrium point for $|\beta |>1$ and a stable limit cycle for $|\beta |<1$. 
Following Crucifix (2012),
the parameters are set as $\alpha =30.0$, $\beta =0.75$, $\gamma =0.4$, and $\tau =36.0$~kyr.  

The dynamical properties of Eq.~(\ref{eq:CSW}) were extensively studied by De~Saeldeleer, Crucifix, and Wieczorek (2013).
Thus, we refer to Eq.~(\ref{eq:CSW}) as the CSW model.

\subsubsection{Paillard--Parrenin model}
Paillard and Parrenin (2004) proposed an oscillator model of glacial cycles involving the Antarctic ice-sheet influence 
on bottom water formation.
The model, which we call the PP04 model, consists of equations for the global ice volume $x$, extent of the Antarctic ice sheet $y$, 
and atmospheric CO$_2$ concentration $z$ as follows:
\begin{equation} 
\begin{split}
\tau _x\dot x& =-pz-qF(t)+r-x,\\
\tau _y\dot y&=x-y,\\
\tau _z\dot z&=\alpha F(t)-\beta x+\gamma H(-w)+\delta -z,\\
w&=ax-by-cI(t)+d,
\end{split} \label{eq:PP04}
\end{equation}
where $F(t)$ is the summer solstice insolation at 65$^\circ $N given by Eq.~(\ref{eq:ins}) with $a=23.58$ W/m$^2$, and 
$I(t)$ is the insolation at 60$^\circ $S on February 21.
The Heaviside function in the original PP04 model is approximated as $H(x)=1/2+\arctan (1000x)/\pi$ for simplicity of the dynamical systems analysis.
The parameters are set as $\tau _x=15$~kyr, $\tau _y=12$~kyr, $\tau _z=5$~kyr, $p=1.3$, $q=0.5$, $r=0.8$, 
$\alpha =0.15$, $\beta =0.5$, $\gamma =0.5$, $\delta =0.4$, $a=0.3$, $b=0.7$, $c=0$, and $d=0.27$.
The choice $c=0$ does not change the timing of terminations; 
it only affects the qualitative agreement with geological records (Paillard and Parrenin 2004).

\subsubsection{Imbrie--Imbrie model}
Imbrie and Imbrie (1980) proposed a piecewise-linear model for the glacial cycles as follows: 
\begin{equation*}
\begin{split}
\tau \dot x&=\left\{
\begin{split}
(1+b)(F(t)-x) \,\,\,\mbox{if}\,\,F(t)\geq x,\\
(1-b)(F(t)-x) \,\,\,\mbox{if}\,\,F(t)<x, 
\end{split}
\right.
\end{split}
\end{equation*}
where $x$ corresponds to a loss in the global ice volume. The variable
$F(t)$ is assumed to be the summer solstice insolation at 65$^\circ $N given by Eq.~(\ref{eq:ins}) with $a=23.58$ W/m$^2$ (cf. Paillard 2001).
The parameters are tuned as $\tau =17$~kyr and $b=0.6$. 
For simplicity of the dynamical systems analysis, 
we deal with a smoothed system,
\begin{equation}
\tau \dot x=[1+b\tanh k(F(t)-x)](F(t)-x). \label{eq:Imbrie}
\end{equation}
If the steepness parameter $k$ is large enough, the solution of Eq.~(\ref{eq:Imbrie}) converges to that of the original model.
We set $k=10$ in this study.

\section{Dynamical systems analysis}
\subsection{Synchronization under the same astronomical forcing}
\begin{figure}[]
\begin{center}
\includegraphics[scale=0.48,angle=0]{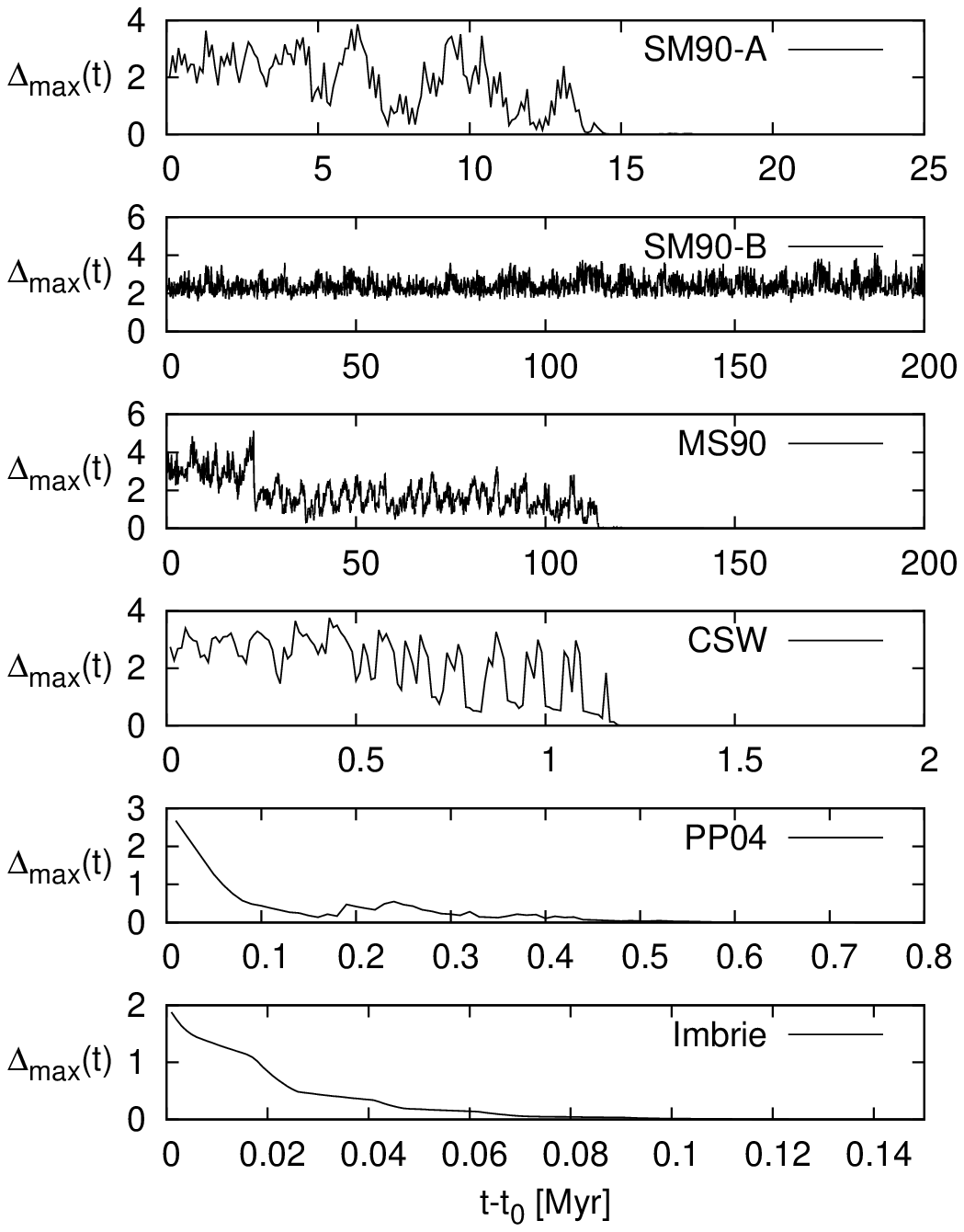}
\caption{Convergence of $10^4$ solutions for each model: 
the maximum distance from the average of the ensemble $\Delta _{\max}(t)=\max _{1\leq i\leq 10^4}\left| \mathbf{x}^{(i)}(t)-\langle \mathbf{x}^{(i)}(t)\rangle\right|$ vs. elapsed time $t-t_0$.
All solutions start from random initial states at time $t_0=-20$~Myr.
Note that the time scale is different for each panel
} \label{fig:as}
\end{center}
\end{figure}

In what follows, $\mathbf{x}(t)$ represents the climate state at time $t$ for each model described in the previous section\footnote{For example, $\mathbf{x}=(x,y,z)^\mathrm{T}$ for the SM90 model, and $\mathbf{x}=x$ for the Imbrie model.}.
First, we check the dependence of solutions on initial climate states under the same astronomical forcing $F(t)$.
We simulate $10^4$ solutions $\mathbf{x}^{(i)}(t)$ ($i=1,\,2,\,...,\,10^4$), starting from random initial states $\mathbf{x}^{(i)}(t_0)$ at time $t_0=-20$~Myr, under the same astronomical forcing $F(t)$.
For these solutions, we calculate the maximum distance from the average of the ensemble, 
$$\Delta _{\max}(t)=\max _{1\leq i\leq 10^4}\left| \mathbf{x}^{(i)}(t)-\langle \mathbf{x}^{(i)}(t)\rangle\right|.$$
We assume that typical solutions converge to the same solution, i.e., {\it synchronize}\footnote{
Following Ramaswamy (1997), we use the term ``synchronization'' in the sense of the convergence of typical orbits start from different initial conditions.},
under the same astronomical forcing
if $\Delta _{\max}(t)\to 0$ as $t\to \infty$.
Figure~\ref{fig:as} shows the behavior of $\Delta _{\max}(t)$ for each model.
In each model except for the SM90-B model, typical solutions synchronize after a certain transient time.
This result implies that the attractor is unique at least except for the SM90-B model.
It should be mentioned that the transient time is longer than the duration of the Quaternary, approximately 2.58~Myr, in the SM90-A and MS90 models (see Fig.~\ref{fig:as}).

\subsection{Simulated ice volume over the last 700~kyr}
\begin{figure}[]
\begin{center}
\includegraphics[scale=0.48,angle=0]{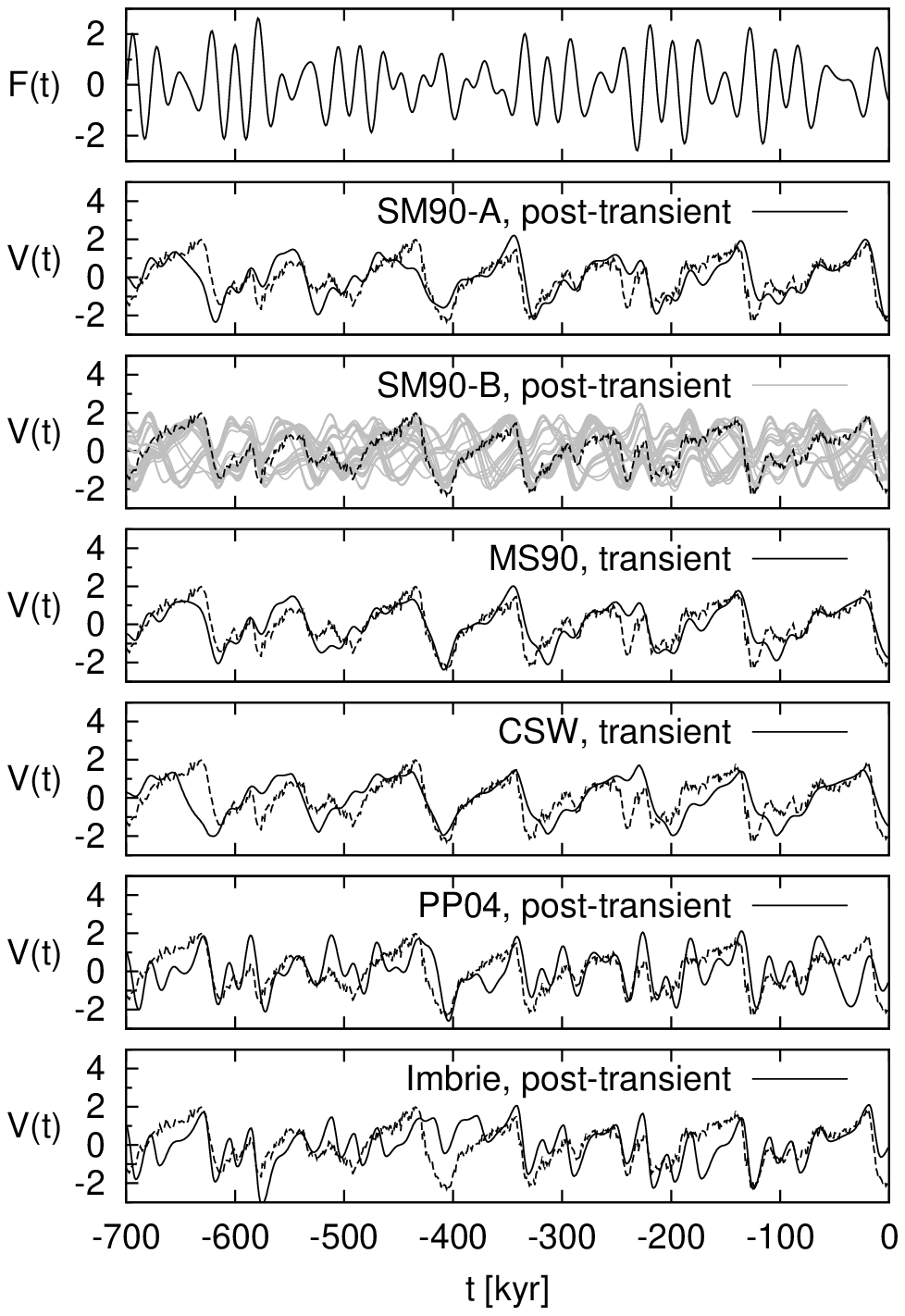}
\caption{{\it Top panel}: Normalized insolation $F(t)$ at 65$^\circ $N on the day of summer solstice. 
{\it Other panels}: Simulated normalized ice volume $V(t)$ ({\it solid lines}) for the models described in Section~3. 
The $\delta ^{18}O$ record of the LR04 stack is also shown for comparison ({\it dashed lines}).
Post-transient solutions are presented for the SM90-A, SM90-B, PP04, and Imbrie models, and transient solutions are presented for the MS90 and CSW models
} \label{fig:compare_700}
\end{center}
\end{figure}
Figure~\ref{fig:compare_700} shows the simulated normalized ice volume $V(t)$ over the last 700~kyr (solid line)
for each model.
The $\delta ^{18}O$ record of the LR04 stack is also shown by a dashed line for comparison (Lisiecki and Raymo 2005).
In the SM90-A, PP04, and Imbrie models, synchronized solutions after each transient period
are well accorded with the $\delta ^{18}O$ record; 
however, in the MS90 and CSW models, synchronized solutions after each transient period are not accorded with the $\delta ^{18}O$ record,
but only some transient solutions agree with the $\delta ^{18}O$ record.
In the SM90-B model, solutions starting from different initial states do not synchronize.
This result indicates that the SM90-B model exhibits chaotic solutions. 

Actually we find coexistence of a quasiperiodic attractor and a chaotic attractor for the SM90-B model.
The quasiperiodic attractor originates in a stable equilibrium of the unforced system 
$(x,y,z)=$(-0.282, 0.277, 0.282).
The basin of attraction of this quasiperiodic attractor is much smaller than that of the chaotic attractor,
and the amplitude of the quasiperiodic solution is also much smaller than that of the chaotic solutions. 
Thus, we do not focus on this quasiperiodic solution.

\subsection{Types of attractors}
We analyze types of attractors using the maximal CLE $\lambda$, the phase sensitivity exponent $\mu$, 
and the parameter sensitivity exponent $\nu $.
As shown in Table~2,
the maximal CLE $\lambda$ is positive for the SM90-B model, and is negative for all the other models.
Thus, the SM90-B model is chaotic, and all the other models are nonchaotic.
This result is consistent with that of convergence experiments. 
\begin{table}
\caption{Characteristic exponents and types of attractors for the different models. 
The asterisk ($\ast$) indicates that the sensitivity exponents are undefined for the SM90-B model.}
\label{tab:2}       
\begin{tabular}{lllll}
\hline\noalign{\smallskip}
Models & $\lambda $ [kyr$^{-1}$] & $\mu $ & $\nu$ & Attractor \\ 
\noalign{\smallskip}\hline\noalign{\smallskip} 
SM90-A & $-0.0014$ & 1.06 & 1.15 & SNA \\
SM90-B & $+0.0020$ & $\ast $ & $\ast $ & Chaotic \\
MS90 & $-0.0005$ & 0.95 & 1.09 & SNA \\
CSW & $-0.016$ & 1.10 & 1.13 & SNA \\
PP04 & $-0.0093$ & 1.09 & 1.08 & SNA\\
Imbrie & $-0.046$ & 0.0 & 0.0 & Quasiperiodic \\
\noalign{\smallskip}\hline 
\end{tabular}
\end{table} 

\begin{figure}[]
\begin{center}
\includegraphics[scale=0.8,angle=0]{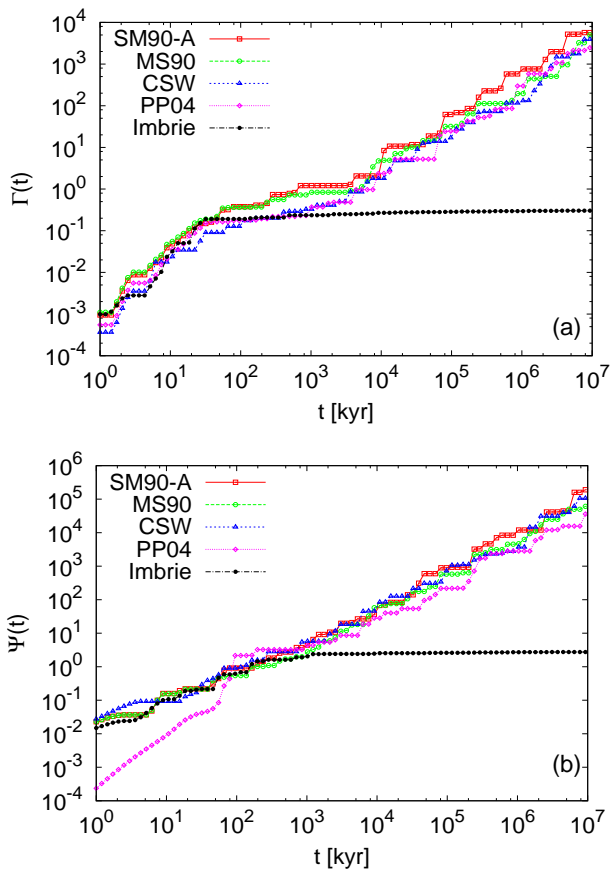}
\caption{{\bf a} Phase sensitivity function $\Gamma (t)$ of $x$ with respect to a phase variable $\theta _4$. {\bf b} Parameter sensitivity function $\Psi (t)$ of $x$ with respect to $u$ for the SM90-A and MS90 models,
parameter $\beta $ for the CSW model, and parameter $b$ for the Imbrie model 
} \label{fig:sm90_ps}
\end{center}
\end{figure}

The next question is whether the attractors are strange or not.
Figure~\ref{fig:sm90_ps}(a) shows the phase sensitivity functions $\Gamma (t)$ of $x$ with respect to phase variable $\theta _4$ for the models with nonchaotic attractors, and
Fig.~\ref{fig:sm90_ps}(b) shows the parameter sensitivity functions $\Psi (t)$ of $x$ with respect to parameter $u$ for the SM90-A and MS90 models, parameter $\beta $ for the CSW model, and parameter $b$ for the Imbrie model.
Here, the functions $\Gamma (t)$ and $\Psi (t)$ are calculated for a set of $20$ initial conditions.
For the Imbrie model, the functions $\Gamma (t)$ and $\Psi (t)$ saturate as time increases,
but for the other models, the functions $\Gamma (t)$ and $\Psi (t)$ 
grow in power-laws with exponents $\mu >0$ and $\nu >0$, respectively.
Table~2 shows the values of $\mu $ and $\nu$, which are estimated from the power-law behaviors for $t>10^4$.
Therefore, the attractors of the SM90-A, MS90, CSW, and PP04 models are SNAs, and the attractor of the Imbrie model is a quasiperiodic attractor.

\subsection{Illustration of attractors by simplification of astronomical forcing}
\begin{figure}[]
\begin{center}
\includegraphics[scale=0.67,angle=0]{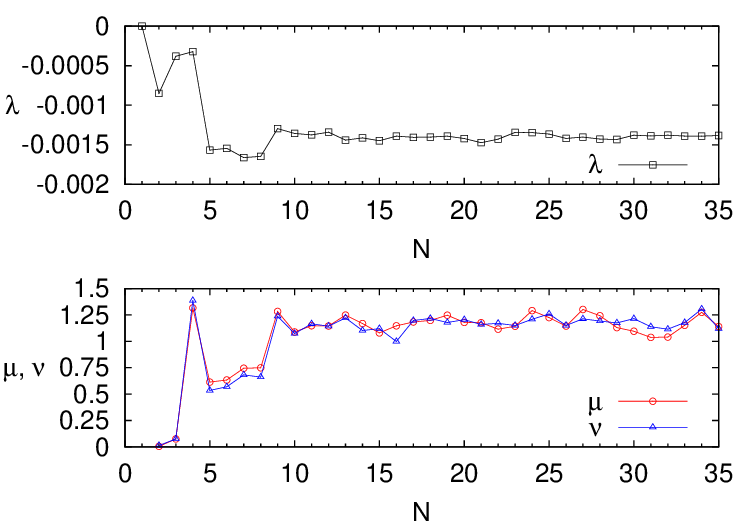}
\caption{Dependence of the characteristic exponents on the number of frequency components $N$ for the case of the SM90-A model. 
{\it Top panel}: The maximal conditional Lyapunov exponent $\lambda$. 
{\it Bottom panel}: The phase sensitivity exponent $\mu$ and the parameter sensitivity exponent $\nu$.
The numerical calculation procedures for these exponents are the same as in Section~4.3.
The attractor for $N=3$ is a quasiperiodic attractor, 
whose tiny but nonzero sensitivity exponents are due to finite computation time 
} \label{fig:Ndependence}
\end{center}
\end{figure}

Since the astronomical forcing $F(t)$ in Eq.~(\ref{eq:ins}) has high degrees of freedom, namely, $N=35$ in our study, 
the attractors in climate models have at least 35 dimensions, 
which are too high-dimensional to be visualized.
To capture the geometric features of these attractors, we illustrate attractors 
for simplified astronomical forcing.
Leaving only the 23.7~kyr ($i=1$) and 19.1~kyr ($i=3$) precession terms and the 41.0~kyr ($i=4$) obliquity term,
we simplify the astronomical forcing $F(t)$ as 
\begin{equation*} 
F(t)=\frac{1}{a_{\{1,3,4\}}}\sum _{i\in \{1,3,4\}} [s_i\sin (\omega _it)+c_i\cos (\omega _it)].
\end{equation*}
These three frequency components constitute 78\% of the original astronomical forcing with respect to standard deviations.
The scale factor is $a_{\{1,3,4\}}=9.08$ W/m$^2$ for the CSW model and $a_{\{1,3,4\}}=18.3$ W/m$^2$ for the other models.
For this simplified forcing\footnote{Other simplified forces using combinations ($i=1,2,3$), ($i=1,2,4$), or ($i=2,3,4$) change the types of attractor for some of the models.}, we obtain the same type of attractor for each model
although the values of characteristic exponents are different from the original values: 
an SNA with $\lambda \approx -0.0004$, $\mu =1.39$, and $\nu =1.43$ for the SM90-A model, 
a chaotic attractor with $\lambda\approx 0.0025$ for the SM90-B model, 
an SNA with $\lambda \approx -0.00012$, $\mu =1.09$, and $\nu =1.35$ for the MS90 model,
an SNA with $\lambda \approx -0.0088$, $\mu =1.13$, and $\nu =1.10$ for the CSW model,
an SNA with $\lambda \approx -0.0089$, $\mu =1.07$, and $\nu =0.94$ for the PP04 model,
and a quasiperiodic attractor with $\lambda \approx -0.046$ and $\mu =\nu =0$ for the Imbrie model.
A number of frequency components are needed to obtain the attractors with almost the same values of characteristic exponents as the original attractors. 
Figure~\ref{fig:Ndependence} shows the dependence of the characteristic exponents on the number of frequency components $N$ in the astronomical forcing $F(t)$ for the case of the SM90-A model,
where the scale factor $a$ is changed depending on $N$ to normalize $F(t)$.
Roughly speaking, $N\geq 9$ is necessary and almost sufficient to approximate the original attractor of the SM90-A model.

\begin{figure*}[]
\begin{center}
\includegraphics[scale=0.11,angle=0]{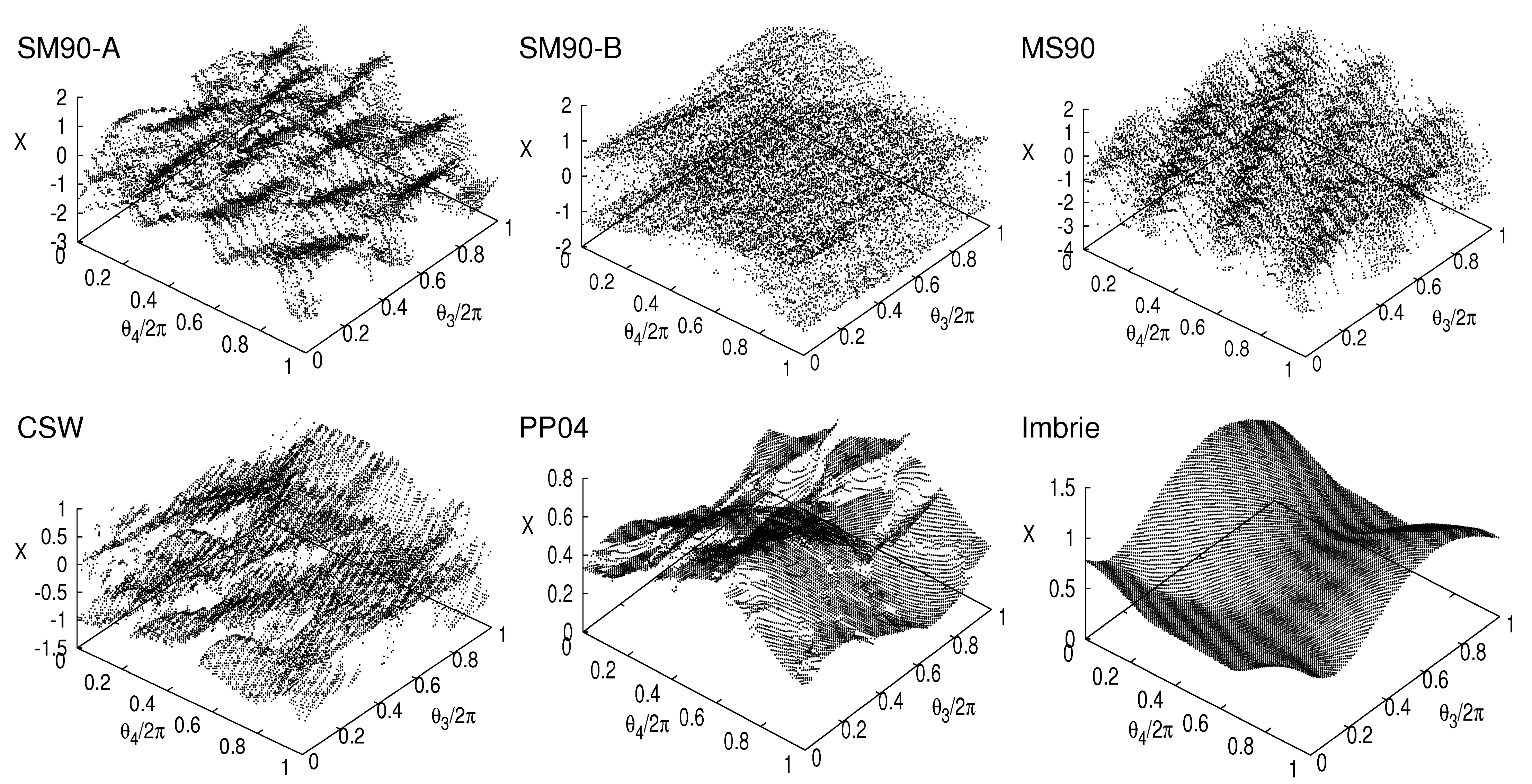} 
\caption{Attractors of each model for the simplified astronomical forcing. Each panel shows the projection of the attractor on the space ($\theta _3/2\pi$, $\theta _4/2\pi$, $x$)
} \label{fig:simple}
\end{center}
\end{figure*}

To visualize the simplified attractors we construct stroboscopic maps.
Observing the system at the moments $t _n=2\pi n/\omega _1+\mbox{const.}$ ($n\in \mathbb{Z}$), where phase $\theta _1(t)$ attains a constant value,
the system reduces to a stroboscopic map ($\theta _3(t_n),$ $\theta _4(t_n)$, $\mathbf{x}(t_n)$) $\mapsto  $ 
($\theta _3(t_{n+1}),$ $\theta _4(t_{n+1})$, $\mathbf{x}(t_{n+1})$).
Figure~\ref{fig:simple} shows the attractors of the simplified models. Each panel presents
the projection of the attractor on the space ($\theta _3(t_n)/2\pi$, $\theta _4(t_n)/2\pi$, $x(t_n)$). 
The SM90-A, MS90, CSW, and PP04 models exhibit SNAs, 
and the dependence of the climate state on the phase of the astronomical forcing is not given by smooth relations,
but constitutes a geometrically strange set.
Moreover, the dependence seems to be discontinuous almost everywhere.
For the Imbrie model, the attractor is quasiperiodic, and the dependence is given by a smooth graph.
For the SM90-B model, the attractor is chaotic, and the dependence constitutes a geometrically strange set,
as in the case of SNAs.

\section{Dynamical properties of the attractors}

\subsection{Asymptotic behavior and transient behavior}
The difference in asymptotic behavior between chaotic and nonchaotic attractors has been shown in the previous section.
If a climate model has a chaotic attractor, 
solutions sensitively depend on initial climate states $\mathbf{x}(t_0)$.
On the other hand, if a climate model has an SNA or a quasiperiodic attractor, 
only one or several synchronized solutions are obtained, regardless of initial climate states $\mathbf{x}(t_0)$, after a certain transient time.

In the previous section, we mentioned that some of the models have a considerably long transient time for synchronization, 
compared to the Quaternary period. 
For such a model, transient behavior is important in practice.
Let us compare the transient dynamics of the SM90-A, SM90-B, and MS90 models.
Since the SM90-B model has the maximal CLE  of 0.002~kyr$^{-1}$, 
we can observe the divergence of nearby orbits within about 500~kyr (the so-called {\it Lyapunov time}).
On the other hand, for the nonchaotic models (SM90-A and MS90), we do not observe such a divergence 
but a slower convergence process of nearby orbits. 
That is, the behavior of nearby orbits is different between the chaotic and nonchaotic attractors. 
However, a chaotic model and a nonchaotic model would not be distinguishable within 1 million years if the maximal CLE of the chaotic model is less than 0.001~kyr$^{-1}$.

\subsection{Sensitivity to dynamical noise} 
Khovanov et al. (2000) stated that SNAs and quasiperiodic attractors differ in sensitivity to perturbations.
We illustrate the difference by simulating the CSW and Imbrie models in the presence of noise.
In the former, we add independent white Gaussian noises $\xi _i(t)$ ($i=x$ or $y$) to the right-hand sides of Eq.~(\ref{eq:CSW}),
respectively, where $\langle \xi _i(t)\rangle =0$, $\langle \xi _i(t)\xi _j(t')\rangle =\sigma ^2 \delta _{ij}\delta(t-t')$, and $\sigma =0.1~\mbox{kyr}^{1/2}$. 
In the latter, we add a white Gaussian noise $\xi _x(t)$ with the same average and standard deviation to the right-hand side of Eq.~(\ref{eq:Imbrie}).
\begin{figure}[]
\begin{center}
\includegraphics[scale=0.47,angle=0]{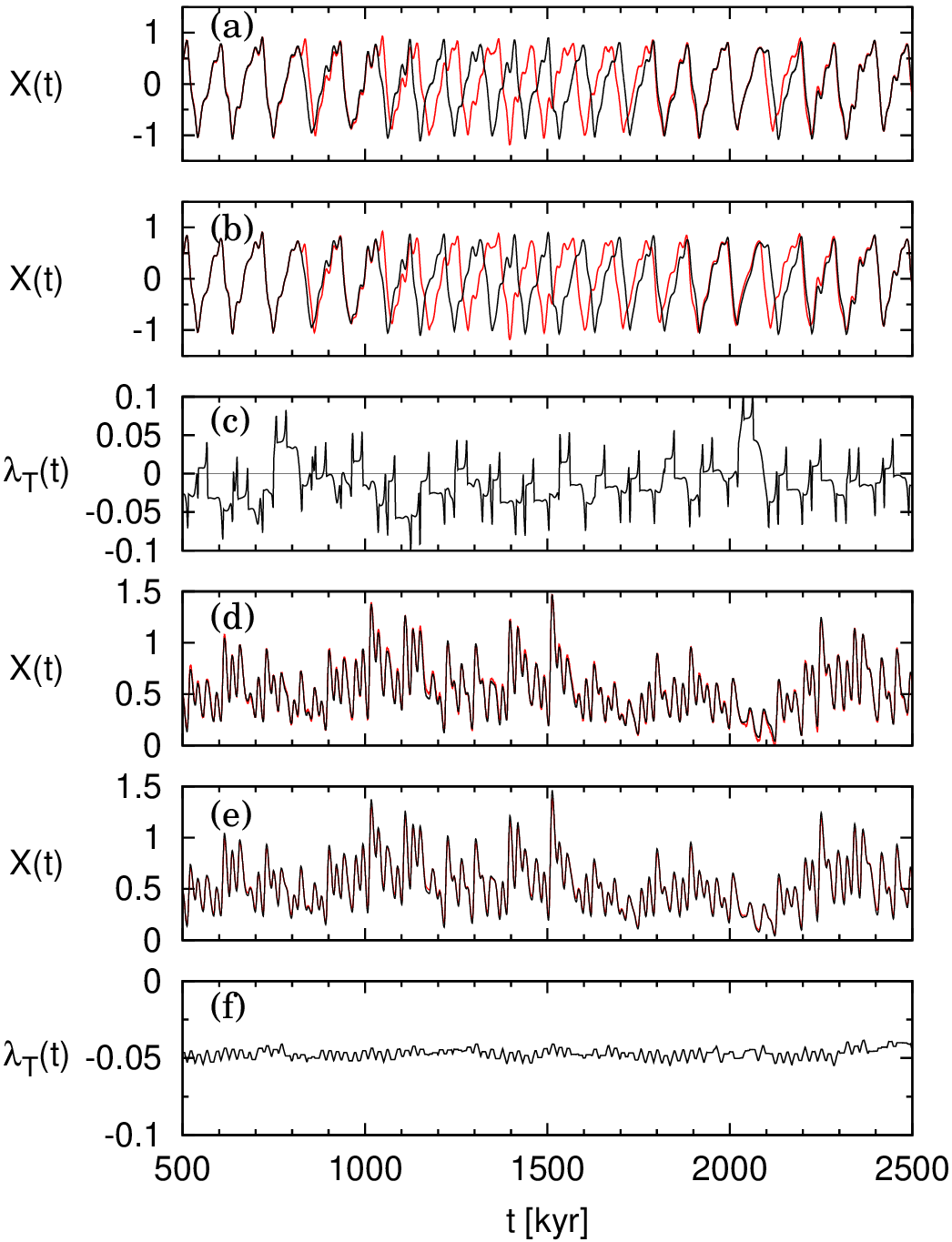}
\caption{Sensitivity of climate models to perturbations: 
{\bf a} Ice volume $x(t)$ simulated by the CSW model with noise ({\it red}) and without noise ({\it black}). 
{\bf b} Ice volume $x(t)$ simulated by the CSW model for $\tau =36.0$~kyr ({\it black}) and for $\tau =36.1$~kyr ({\it red}).
The other parameter values are the same as in Section 3.2.3. 
{\bf c} The finite-time CLE $\lambda _T(t)$ of the unperturbed trajectory of the CSW model in panels {\bf a} and {\bf b}. $T=70$~kyr.
{\bf d} Loss of ice volume $x(t)$ simulated by the Imbrie model with noise ({\it black}) and without noise ({\it red}). 
{\bf e} Loss of ice volume $x(t)$ simulated by the Imbrie model for $\tau =17$~kyr ({\it black}) and for $\tau =19$~kyr ({\it red}).
The other parameter values are the same as in Section 3.2.5.  
{\bf f} The finite-time CLE $\lambda _T(t)$ of the unperturbed trajectory of the Imbrie model in panels {\bf d} and {\bf e}. $T=70$~kyr
} \label{fig:noise}
\end{center}
\end{figure}

Figure~\ref{fig:noise}(a) shows two trajectories of the CSW model simulated with and without noises, 
and Fig.~\ref{fig:noise}(d) shows those of the Imbrie model.
In each model, perturbed and unperturbed trajectories start from the same initial climate state $\mathbf{x}(t_0)$ at $t_0=-1.5$~Myr,
and the noise is added to each system from $t=0$~kyr.
The CSW model is more sensitive to noise than the Imbrie model because of the local instability of SNA.
To characterize the local instability of attractors, let us define a finite-time CLE
$$\lambda _T(t)=\frac{1}{T}\ln \frac{|\delta \mathbf{x}(t+T)|}{|\delta \mathbf{x}(t)|}.$$
The finite-time CLE $\lambda _T(t)$ gives the rate of exponential divergence of nearby orbits 
during the time interval $[t,t+T]$.
Note that $\lambda _T(t)$ converges to the maximal CLE $\lambda$ as $T\to \infty$.
For SNAs, the finite-time CLE $\lambda _T(t)$ can be positive even for large $T$ with nonzero probability (Pikovsky and Feudel 1995).
Thus, the orbits of SNAs are sensitive to perturbations in time intervals with $\lambda _T(t)>0$.
Figures~\ref{fig:noise}(a) and \ref{fig:noise}(c) show that the perturbed trajectory occasionally deviates 
after the finite-time CLE $\lambda _T(t)$ becomes positive, where $T=70$~kyr.
This local instability may be attributed to a repellor embedded in SNA as mentioned in Section~2.
Such a temporary loss of synchronization has been shown also in another model of glacial cycles (Crucifix 2011).
The frequency and the duration of desynchronization increase as the magnitude of noise $\sigma$ increases.
Therefore, the orbits of SNAs and those of chaotic attractors might be indistinguishable under large dynamical noise.

On the other hand, for quasiperiodic attractors, the finite-time CLE $\lambda _T(t)$ takes only negative values for large $T$. 
As a result, the orbits of quasiperiodic attractors are robust to perturbations in comparison with those of SNAs.
Figures~\ref{fig:noise}(d) shows that the Imbrie model is robust to the same magnitude of noise that can cause larger deviations of trajectories in the CSW model. 
For the Imbrie model, the finite-time CLE $\lambda _T(t)$ with $T=70$~kyr is always negative as shown in Fig.~\ref{fig:noise}(f). 

If a climate model has a nonchaotic attractor in the absence of noise, there can be a relation from the phase of the astronomical forcing to the climate state.
The existence of such an intact phase relation is assumed when we refine a time scale of a paleoclimate record based on astronomical cycles (the so-called {\it orbital tuning}).
However, the above results suggest that such a phase relation is rather robust for quasiperiodic attractors but not robust for SNAs.

\section{Discussion}
Let us mention an implication of this study for paleoclimate modeling. The notion of SNAs will be helpful for understanding apparently-contradictory properties of conceptual models of glacial cycles. In a number of previous studies, it is reported that a model of glacial cycles is sensitive to subtle changes in parameter values although the model is insensitive to initial conditions (for example, Weertman 1978; Oerlemans 1982; Paillard 2001; Ganopolski and Calov 2011; Crucifix 2013). 
In some models, such parameter sensitivity may be attributed to thresholds with discontinuity (Paillard 2001). 
However, parameter sensitivity can arise in continuous models when they have SNAs.
Figure~\ref{fig:noise}(b) shows strong parameter sensitivity of the CSW model with SNA.
The black line represents the original solution for $\tau =36.0$~kyr,
and the red line represents the solution for $\tau =36.1$~kyr (the parameter change is only 2.8\% of the original value).
Figure~\ref{fig:noise}(e) shows weak parameter sensitivity of the Imbrie model with quasiperiodic attractor.
The black line represents the original solution for $\tau =17$~kyr,
and the red line represents the solution for $\tau =19$~kyr (the parameter change is 10.5\% of the original value). 

In a recent paper, Crucifix (2013) investigates the causes of parameter sensitivity inherent in various conceptual models of glacial cycles, which can make the prediction of ice ages difficult (see also Added note). Crucifix (2013) attributes the parameter sensitivity to two causes: complicated bifurcation structures with numerous {\it frequency-locking tongues} and abrupt changes of attractors with respect to subtle changes in parameter values. 
We conjecture that such abrupt changes of attractors are due to a property of SNAs with strange geometry. 
The validation of this conjecture needs detailed bifurcation analyses and will be an important future study.

\section{Conclusion} 
We analyzed five conceptual models of glacial cycles in the framework of the dynamical systems theory.
The SM90-A, MS90, CSW, and PP04 models exhibit a strange nonchaotic attractor (SNA), 
and the dependence of the climate state on the phase of the astronomical forcing is not given by smooth relations,
but constitutes a geometrically strange set.
The SM90-B model exhibits a chaotic attractor, and 
the dependence of the climate state on the phase of the astronomical forcing is not given by smooth relations,
but constitutes a geometrically strange set.
The Imbrie model exhibits a quasiperiodic attractor, and 
the dependence of the climate state on the phase of the astronomical forcing is given by a smooth graph.
These results suggest that SNA is a candidate for representing the dynamics of glacial cycles,
in addition to well-known quasiperiodicity and chaos\footnote{Chaos in conceptual models of glacial cycles has often been mentioned in the literature (e.g. Nicolis 1987; Saltzman and Verbitsky 1992, 1993; Ghil 1994; Huybers 2009). 
Rial (2004) proposed an idea that the climate operates at the edge between chaos and order at orbital
and millennial scales.
Brindley and Kapitaniak (1992) specified an {\it inhibition of chaotic behavior} 
due to the quasiperiodicity of forcing.
In their article, the inhibition of chaotic behavior refers to the appearance of statistical periodicity for the ensemble of orbits,
and it is different from the nonchaoticity based on nonpositive Lyapunov exponents in this study.}.
Since SNAs have repellors in themselves, the orbits on SNAs are locally unstable.   
Therefore, if the real climate system has an SNA, the orbit will be temporarily sensitive to perturbations 
when the finite-time conditional Lyapunov exponent is positive.

The analysis in this paper was restricted to spatially zero-dimensional models 
that can be approximated by smooth dynamical systems. 
A {\it hybrid dynamical system} is a system that has both continuous and discrete dynamics (Aihara and Suzuki 2010).
A number of ice age models are expressed as hybrid dynamical systems, more precisely, {\it hybrid automata}\footnote{A simple example of hybrid automata is the temperature-thermostat system,
where a continuous variable is the temperature and a discrete variable is the status of the heater $\{$on, off$\}$.}
(van der Schaft and Schumacher 2000).
For example, the models by Paillard (1998), Parrenin and Paillard (2003), and Ashkenazy and Tziperman (2004) 
are hybrid automata.
The Imbrie and PP04 models are also hybrid dynamical systems 
but can be approximated by smooth dynamical systems because they are especially {\it piecewise affine systems}.
The description of the dynamics appearing in hybrid automata models or 
in more complicated models with spacial dimensions is still unclear.   
This problem should be studied in the future. 

\section*{Added note}
After the submission of our article, we came to know a discussion paper submitted at about the same time, which mentions SNAs in an ice age model (Crucifix M, Why could ice age be unpredictable? Clim Past Discuss 9:1053-1098, 2013). The author demonstrates the appearance of SNAs in the CSW model for simplified forcing with two frequencies, which is different from the forcing used in our article. 
The verification method of SNAs is also different between two articles. 
We use the maximal conditional Lyapunov exponent and the sensitivity exponents. 
On the other hand, Crucifix (2013) uses the pullback section method described in De~Saedeleer et al. (2013). 
Thus, the results obtained by Crucifix and by us are distinguishable and complementary to each other. 
The discussion in Section~6 is motivated by the paper (Crucifix 2013).

\begin{acknowledgements}
We thank Dr.~Hiroki Takahasi and Dr.~Naoya Fujiwara for their valuable comments.
This research is supported by the Aihara Innovative Mathematical
Modelling Project, the Japan Society for the Promotion of Science
(JSPS) through the ``Funding Program for World-Leading Innovative R\&D
on Science and Technology (FIRST Program),'' initiated by the Council
for Science and Technology Policy (CSTP).
\end{acknowledgements}

\section*{Appendix 1: Parameter values for the insolation formula}
The parameter values for the insolation formula~(\ref{eq:ins}) given by De~Saedeleer et al. (2013) are listed in Table~\ref{tab:3} to make the paper self-contained.
The frequencies and coefficients are arranged in descending order of their power $s_i^2+c_i^2$.
\begin{table*}
\begin{center}
\caption{Parameter values for the insolation formula of De~Saedeleer et al. (2013)}
\label{tab:3}       
\begin{tabular}{llll}
\hline\noalign{\smallskip}
Index $i$ & $\omega _i$ [rad/kyr] & $s_i$ [W/m$^2$] & $c_i$ [W/m$^2$]\\ 
\noalign{\smallskip}\hline\noalign{\smallskip} 
1 & 0.264933601588513 & -15.549049332290400 & -9.704062871105320 \\
2 & 0.280151350350945 & 15.431955636170100 & 4.752472711315250 \\
3 & 0.331110950251899 & 9.099224935273400 & -10.611524488739001 \\
4 & 0.153249478547167 & -11.228737681512399 & 3.516820752112410 \\
5 & 0.328024059125949 & -7.870653840136690 & 6.615442460635030\\
6 & 0.326211762560183 & 0.813786144754451 & -4.526414080992460 \\
7 & 0.158148666238883 & -3.824993714675400 & -0.761851750263805 \\
8 & 0.269742342439881 & 0.069044850431486 & -3.316392609695580 \\
9 & 0.117190147169570 & 2.288148059560660 & 1.802337026846230 \\
10 & 0.332923246817665 & 1.440507707859670 & 1.063392860501200 \\
11 & 0.150162587421217 & 1.549041763533020 & -0.088394191276982 \\
12 & 0.217333905941751 & 0.380973541305497 & -1.463017119992100 \\
13 & 0.155061775112933 & -1.297700819564400 & -0.635152963728496 \\
14 & 0.371638925683567 & 0.925324276580528 & -1.020667586721540 \\
15 & 0.324937167999999 & -0.576082669762308 & 1.186695727393380 \\
16 & 0.275366617473065 & 0.997628846513796 & -0.362906496840039 \\
17 & 0.211709630908568 & -0.810768209286259 & -0.577980646565494 \\
18 & 0.156336369673117 & -0.918358442095885 & 0.196083726889428 \\
19 & 0.334197841377850 & 0.346906064369828 & -0.648189701487285 \\
20 & 0.259396912994958 & 0.339477750517033 & -0.560509461538342 \\
21 & 0.323124871434233 & -0.378637986107629 & 0.527217891742183 \\
22 & 0.148350290855451 & 0.256895610735773 & -0.524697312305024 \\
23 & 0.111684123041346 & -0.428006728186239 & 0.357006342316690 \\
24 & 0.336010137943616 & -0.421809264016129 & 0.324327509437558 \\
25 & 0.274551083291250 & -0.441772417569753 & 0.289576210423804 \\
26 & 0.126901871803777 & 0.257509918217341 & -0.377639794223366 \\
27 & 0.177861471704732 & -0.161827722328271 & -0.362683869407858\\
28 & 0.206924898030688 & -0.335783913402678 & -0.019479215012864 \\
29 & 0.212525165090383 & 0.267659228540196 & 0.128915417116900 \\
30 & 0.004899187691716 & -0.160479848721994 & 0.059407796893426 \\
31 & 0.418183080135680 & -0.018488406464501 & 0.109632390175297 \\
32 & 0.433400828898112 & -0.004919921945456 & -0.106148873639336 \\
33 & 0.229992875969202 & 0.069618973318896 & 0.074623171406128 \\
34 & 0.311398144786051 & 0.013835372762118 & 0.030473666884042\\
35 & 0.306498957094334 & 0.024734974816962 & 0.014046439534097 \\
\noalign{\smallskip}\hline 
\end{tabular}\\
\end{center}
\end{table*}


\end{document}